\documentclass[aps,pra,reprint,amsmath,amssymb,showpacs]{revtex4-1}

\usepackage{subfigure}
\usepackage[breaklinks=true]{hyperref}
\usepackage{graphicx}
\newcommand{\bra}[1]{\langle #1|}
\newcommand{\ket}[1]{|#1\rangle}
\newcommand{\braket}[2]{\langle #1|#2\rangle}
\newcommand{\dd}{\textrm{d}}
\newcommand{\ii}{\textrm{i}}
\newcommand{\ee}{\textrm{e}}

\begin{document}

\title{Quantum optical reconstruction scheme using weak values}

\author{Joachim Fischbach}
\email{joachim.fischbach@uni-ulm.de}

\author{Matthias Freyberger}

\affiliation{Institut f\"ur Quantenphysik, Universit\"at Ulm, D-89069 Ulm,
 Germany}

\date{November 9, 2012}

\begin{abstract}
A quantum state contains the maximal amount of information available for a
given quantum system. In this paper we use weak-value expressions to
reconstruct quantum states of continuous-variable systems in the quantum optical
domain. The role played by postselecting measured data will be particularly
emphasized in the proposed setup, which is based on an interferometer just using
simple homodyne detection.
\end{abstract}


\maketitle

\section{Introduction}

Since the launch of the idea and formalism by Aharonov, Albert, and Vaidman
\cite{aav, cl:aharonovvaidman} a considerable amount of research has
focused on weak measurements. Basic questions behind the concept have been
clarified in a series of papers \cite{dss, ge:resch, ge:johansen, ge:oreshkov,
cl:aharonovbotero, jozsa, mitchisonjozsa}, and it was possible to extend the
original proposal in various directions \cite{ge:dilorenzo, ge:wu,
ge:parksgray, dresseljordan, ge:nakamura}. It is fascinating to see how the
weak-value formalism touches very different fields like phase-space physics
\cite{pslr, psdg}, quantum trajectories \cite{ge:wiseman, wiseman},
contextuality \cite{dresselagarwal}, and quantum cloning \cite{ge:hofmann}.

In parallel to these theoretical developments the experimental realization
\cite{ce:ritchie} of weak measurements has recently found a rich territory of
applications \cite{ce:pryde, ce:hostenkwiat, ce:dixonstarling}. Even the most
fundamental debates about the realism of quantum observables
\cite{eop:lundeensteinberg, eop:yokota, eop:Goggin} and the role of trajectories
in double-slit experiments \cite{wp:mirlundeen, wp:kocsis} can now be attacked
with the assistance of weak measurements carried out in the laboratory.

In the present paper we concentrate on yet another crosslink:
Lately quantum state reconstruction \cite{leonhardt, paris2004quantum} using
weak measurements has become a topic of great interest \cite{lundeennature2011,
haapasalo, lundeenbamber2012, aqsr:degosson, Kalev}. 
Several works analyze the aspect of a weak interaction \cite{aqsr:silberfarb,
aqsr:smith, aqsr:riofrio} in non-standard quantum state estimation. Obviously
there is a special role played by postselection in creating weak values of
observables. This thought is also pursued in Ref.\ \cite{lundeennature2011} 
where a quantum wave function has been reconstructed using the special form of
weak values. Later this procedure was generalized for mixed quantum states
\cite{lundeenbamber2012}, drawing connections between the structure of weak
values and certain phase-space distributions \cite{kirkwood, dirac, rihaczek}. A
similar connection was also established in Refs.\ \cite{johansenrwvwwm} and
\cite{johansenqtspm} and later used to build yet another reconstruction scheme
\cite{johansenqmsmwamc, Kalev}.

In this article we first recall the special form of weak values used to derive
reconstruction relations. Our focus then lies in establishing a direct
connection between a quantum optical implementation and these reconstruction
relations. The main contribution of this work therefore consists of working
out a realizable quantum optical scheme that fully exploits the form of weak
values to reconstruct a continuous-variable state of a single mode of the
quantized electromagnetic field.

In Sec.\ \ref{sec:wm} we will recapitulate the formalism of
weak measurements. Section \ref{sec:recrel} provides the reconstruction
relations for quantum states of continuous-variable systems. In Sec.\
\ref{sec:qopimp} we show how to realize the method with a quantum optical setup.
Finally, in Sec.\ \ref{sec:sim} we simulate the optical experiment for a special
quantum state of light and discuss to which class of states the reconstruction
scheme can be applied.

\section{Weak measurements \label{sec:wm}}

In this section we will briefly recall the concepts of weak measurements
\cite{aav} to introduce our notation and to make the paper self-consistent. An
appropriate scaling allows us to choose dimensionless quantities everywhere.

\begin{figure}[b]
 \includegraphics{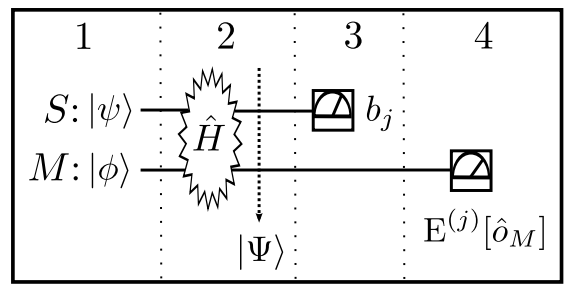}
 \caption{Basic setup of a weak measurement which can be applied to reconstruct
the unknown state $\ket{\psi}$ of a system $S$ with the help of a meter $M$
initially prepared in a specific state $\ket{\phi}$. After system and meter have
been separately defined in the first step, they interact weakly in the second
step. This leaves $S$ and $M$ in the joint state $\ket{\Psi}$. After a suited
projective measurement with outcome $b_j$ in the third step, the postselected
expectation value $\textrm{E}^{(j)}[\hat{o}_M]$ of a convenient meter observable
$\hat{o}_M$ can be obtained in the fourth step. We demonstrate in
Sec.\ \ref{sec:recrel} that the probability distribution of finding the eigenvalue
$b_j$ and the expectation value $\textrm{E}^{(j)}$ are sufficient to reconstruct
the state $\ket{\psi}$.
\label{fig:wm}}
\end{figure}

A weak measurement can be divided into four successive steps as depicted in
Fig.\ \ref{fig:wm}.
In the first step the system $S$ is given in the unknown state $\ket{\psi}$
and the meter $M$ is prepared in the state $\ket{\phi}$
which we assume to be the Gaussian \footnote{This is actually the simplest
assumption which allows us to demonstrate the quantum state reconstruction
scheme in the following sections. Moreover, it is a rather natural assumption
to prepare a meter in a minimal-uncertainty state.}
\begin{equation}
 \ket{\phi} = \frac{1}{\pi^{1/4}} \int\limits_{-\infty}^{\infty} \dd x \
\ee^{-\frac{x^2}{2}} \ket{x}  \label{eq:specmdstate}
\end{equation}
in the scaled position $x$ with vanishing first moments.
In the second step system and meter interact. At this point the
interaction will be simply modeled by the von Neumann Hamiltonian \cite{vnm}
\begin{equation}
 \hat{H} = g \delta(t-t_0) \hat{A}_S \hat{p}_M , \label{eq:vNMHamiltonian}
\end{equation}
where $g$ is the interaction strength and $\hat{A}_S$ represents a certain
observable of the system which is coupled to the momentum operator $\hat{p}_M$
of the meter. The interaction happens instantaneously at time $t_0$ and is weak,
i.e.\ $g$ is small. In Sec.\ \ref{sec:qopimp} all these model-like assumptions
are shown to be implementable in a quantum optical setup for state
reconstruction.

A postselective measurement is then performed in the third and fourth step.
Postselection begins by making a projective measurement of an observable
$\hat{B}_S$ of the system. We obtain one of its eigenvalues $b_j$ with
probability
\begin{equation}
  p(b_j) = \bra{\Psi} \left( \ket{b_j}_S\bra{b_j} \otimes \hat{\openone}_M
\right) \ket{\Psi} \label{eq:psprob} ,
\end{equation}
where $\ket{\Psi} \equiv \exp[ -\ii g \hat{A}_S \hat{p}_M ]\ket{\psi}\ket{\phi}$
denotes the joint state of system and meter after the interaction.

The fourth and final step consists of measuring a suitable observable
$\hat{o}_M$ of the meter, given that we have found the eigenvalue $b_j$ before.
The corresponding postselected expectation value 
\begin{equation}
 \textrm{E}^{(j)}[\hat{o}_M] = \frac{1}{p(b_j)} \bra{\Psi} \left(
\ket{b_j}_S\bra{b_j} \otimes \hat{o}_M \right) \ket{\Psi} \label{eq:exactexpval}
\end{equation}
can now be approximated by keeping all terms up to linear order in the coupling
$g$. With the notation $[,]$ for the commutator and $\{,\}$
for the anticommutator, this yields \cite{jozsa}
\begin{align}
 \textrm{E}^{(j)}[\hat{o}_M] =&  \ii g \textrm{Re}[A_w^{(j)}] \bra{\phi}
[\hat{p}_M,\hat{o}_M] \ket{\phi} + \nonumber \\ & + g \textrm{Im}[A_w^{(j)}]
\bra{\phi}\{\hat{p}_M,\hat{o}_M\} \ket{\phi} \label{eq:fullexpval} , 
\end{align}
where we have used the vanishing first moments of $\ket{\phi}$,
Eq.\ (\ref{eq:specmdstate}), and the definition of the weak value \cite{aav}
\begin{equation}
 A_w^{(j)} \equiv \frac{\bra{b_j}\hat{A}_S\ket{\psi}}{\braket{b_j}{\psi}}
\label{eq:defwv} . 
\end{equation}
The use of weak measurements for quantum state reconstruction
\cite{lundeennature2011,lundeenbamber2012,aqsr:degosson,johansenqmsmwamc,Kalev}
relies on the special form of this definition. Notably, it resembles expressions
as they are used in interferometric state reconstruction algorithms
\cite{freyberger}. Before using this special form of the definition of the weak
value to derive reconstruction relations in Sec.\ \ref{sec:recrel}, we finally
look at the case where the observable $\hat{o}_M$ to be measured in step 4 is
the scaled position $\hat{x}_M$ of the meter. Using the relations
\begin{align}
 \bra{\phi}\{\hat{p}_M,\hat{x}_M\} \ket{\phi} = 0 \qquad  \text{and} \qquad
 [\hat{p}_M,\hat{x}_M]= - \ii \label{eq:comrel}
\end{align}
we can simplify the expectation value Eq.\ (\ref{eq:fullexpval}) and arrive at
\begin{align}
 \textrm{E}^{(j)}[\hat{x}_M] =& g \textrm{Re}\left[A_w^{(j)} \right] .
\label{eq:specexpval}
\end{align}
So, for this special choice of meter observables, the real part of the weak
value $A_w^{(j)}$ is a measurable quantity.

\section{Reconstruction relations \label{sec:recrel}}

In this section we will present the very heart of our reconstruction scheme.
It turns out that the real part of a weak value as in Eq.\ (\ref{eq:specexpval})
and the postselection probability Eq.\ (\ref{eq:psprob}) can be used to
reconstruct an unknown state  $\ket{\psi}$ of the system. We will focus on the
case leading to reconstruction relations for the momentum representation 
\begin{equation}
 \braket{P}{\psi} \equiv \psi(P) \equiv |\psi(P)|\ee^{\ii \varphi(P)} .
\label{eq:polardecomp}
\end{equation}
Note that this means to reconstruct the modulus $|\psi(P)|$ and the phase
$\varphi(P)$. This also means that the relations which we will present are not
directly applicable to the case of a mixed state.

We start by regarding a weak measurement of the position observable
$\hat{A}_S=\hat{X}_S$ and a projective measurement of the
conjugate variable, the momentum $\hat{B}_S=\hat{P}_S$, in the postselection
step. 
If we insert these special choices into the definition of the weak value,
Eq.\ (\ref{eq:defwv}), we get
\begin{equation}
  X_w^{(P)} \equiv \frac{\bra{P}\hat{X}_S \ket{\psi}}{\braket{P}{\psi}} = 
\frac{\ii \frac{\dd}{\dd P} \psi(P) }{\psi(P)} ,
\end{equation}
where we have also used the momentum representation of the position observable
$\hat{X}_S $.
Finally, rewriting this with the polar decomposition
Eq.\ (\ref{eq:polardecomp}) results in
\begin{equation}
  X_w^{(P)} = \ii \ \frac{\frac{\dd}{\dd P}|\psi(P)|  }{|\psi(P)|} - \frac{\dd
\varphi(P)}{\dd P} . \label{eq:leavenswvrel}
\end{equation}
An analogous relation was first mentioned in the context of Bohmian mechanics
and its connection to weak values \cite{leavens, wiseman, dresseljordan}. We
will utilize it to derive a reconstruction relation which then turns out to be
particularly suited for the quantum optical implementation in the next section.
By taking the real part of Eq.\ (\ref{eq:leavenswvrel}) we find in particular
that
\begin{equation}
  \textrm{Re}\left[ X_w^{(P)} \right] = - \frac{\dd \varphi(P)}{\dd P} .
\label{eq:rewvphase}
\end{equation}
If we now again use the special choice of observables $\hat{A}_S=\hat{X}_S$ and
$\hat{B}_S=\hat{P}_S$ in Eq.\ (\ref{eq:specexpval}) and combine the result with
Eq.\ (\ref{eq:rewvphase}), we arrive at
\begin{equation}
 \frac{\dd \varphi(P)}{\dd P} = - \frac{1}{g} \textrm{E}^{(P)}[\hat{x}_M] .
\end{equation}
This leads directly to the reconstruction relation for the phase,
\begin{equation}
 \varphi(P) = - \frac{1}{g} \int\limits_{0}^{P} \dd P'\
\textrm{E}^{(P')}[\hat{x}_M]  , \label{eq:recrelphase}
\end{equation}
in which the lower integral bound is arbitrary, since any global phase of the
state $\ket{\psi}$ cannot be detected.

To complete the state reconstruction we still need the absolute value
$|\psi(P)|$ of the system state in momentum representation. Hence we
calculate the probability Eq.\ (\ref{eq:psprob}) of finding the eigenvalue
$P$ in the projective measurement on the system. When we again retain only terms
up to linear order in $g$, we arrive at
\begin{equation}
 p(P) =  \bra{\Psi} \left( \ket{P}_S\bra{P} \otimes \hat{\openone}_M \right)
\ket{\Psi} = |\psi(P)|^2 ,  \label{eq:psprobapprox}
\end{equation}
keeping in mind that the first moments of the meter state
$\ket{\phi}$, Eq.\ (\ref{eq:specmdstate}), vanish.
Thus the reconstruction relation for the absolute value reads
\begin{equation}
 |\psi(P)| = \sqrt{p(P)} . \label{eq:recrelabs}
\end{equation}
Note that the information on the modulus $|\psi(P)|$ just comes from simple
projective measurements of momentum performed in step 3 of the scheme shown
in Fig.\ \ref{fig:wm}. The corresponding information on the phase $\varphi(P)$,
according to Eq.\ (\ref{eq:recrelphase}), can be extracted only from a set of
postselected data, i.e., a continuum of expectation values ordered with respect
to the momentum of the system.

\section{Quantum optical implementation \label{sec:qopimp}}

The reconstruction relations introduced in the previous section rely on the
special form of the interaction Hamiltonian Eq.\ (\ref{eq:vNMHamiltonian}).
So it is a crucial question to ask where in nature the momentum observable of 
one subsystem, here $\hat{p}_M$ of the meter $M$, is coupled to the position
observable of another subsystem, here $\hat{X}_S$ of the unknown system $S$.
In the next section we recall a straightforward optical realization for which
the derived reconstruction relations hold. For another optical version with
a very different aim, see Ref.\ \cite{oimp:wu}.

\subsection{Beam splitter interaction}

In linear quantum optics \cite{schleichQO} a beam splitter simply couples two
modes of light; see Fig.\ \ref{fig:qoimp}. One mode represents the system $S$
prepared in the state $\ket{\psi}$ and the second mode will be the meter in
state $\ket{\phi}$. The joint state after the beam splitter reads
\cite{leonhardt, schleichQO} 
\begin{equation}
 \ket{\Psi} = \exp \left[- \ii \ \theta \left( \hat{X}_S \hat{p}_M - \hat{P}_S
\hat{x}_M \right) \right] \ket{\psi} \ket{\phi} , \label{eq:bsint}
\end{equation}
where $\theta$ is the parameter that determines reflection and transmission of
the device \footnote{In more detail this means that a reflection coefficient 
$\sin \theta$ quantifies how strongly the meter mode couples to the system mode
and vice versa. Hence for $\theta \ll 1$ we obtain a beam splitter with very low
reflection, which corresponds to the weak-coupling regime. On the other hand,
the value $\theta=\pi/4$ marks a 50:50 beam splitter with balanced reflection
and transmission.}. Here the observables
\begin{equation}
 \hat{X}_S = \frac{1}{\sqrt{2}} \left( \hat{a}_S + \hat{a}_S^{\dagger} \right)
\ \text{and} \ \hat{P}_S = \frac{1}{\ii \sqrt{2}}
\left( \hat{a}_S - \hat{a}_S^{\dagger} \right)
\end{equation}
are the position and momentum quadrature observables \cite{schleichQO} of the
system mode, which are defined by the usual creation and annihilation operators
$\hat{a}^{\dagger}_S$ and $\hat{a}_S$. Analogous relations hold for the
quadratures $\hat{x}_M$ and $\hat{p}_M$ of the meter mode. Hence we see that
the beam splitter interaction closely resembles the interaction modeled by the
Hamiltonian of Eq.\ (\ref{eq:vNMHamiltonian}). The additional term, determined
by the product $\hat{P}_S \hat{x}_M$, will actually not affect the postselected
expectation value Eq.\ (\ref{eq:specexpval}).

\subsection{Implementation}

We can now specify the calculations in Sec.\ \ref{sec:wm} using the beam
splitter interaction given by Eq.\ (\ref{eq:bsint}).
The state $\ket{\psi}$ of the system mode is still arbitrary; the
special state of the meter mode is given in Eq.\ (\ref{eq:specmdstate}),
where $\ket{x}$ is now a position quadrature eigenstate. Thus Eq.\
(\ref{eq:specmdstate}) is just the $x$ representation of the vacuum state
$\ket{\phi}=\ket{0}$.

\begin{figure}[t]
 \includegraphics{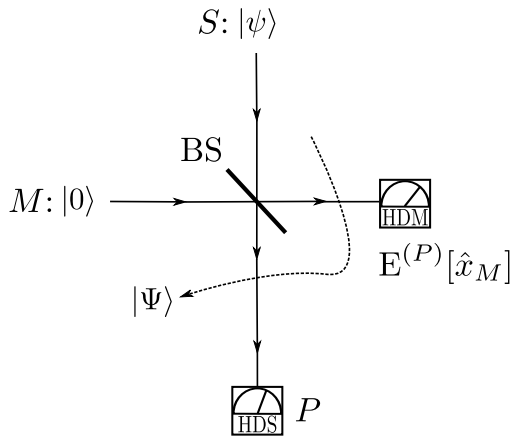}
 \caption{Schematic drawing of a possible quantum optical implementation.
The system mode $S$ and meter mode $M$ are prepared in their respective states and
get coupled by a beam splitter (BS) leaving them in the joint state
$\ket{\Psi}$. Afterwards, using homodyne detectors HDS and HDM, momentum and
position quadratures are measured. Based on the momentum quadrature readings on
HDS we can postselect the expectation values of the position quadrature at HDM.
\label{fig:qoimp}}
\end{figure}

The first step of the postselection is now a projective measurement of the
momentum quadrature observable $\hat{B}_S=\hat{P}_S$, as depicted in
Fig.\ \ref{fig:qoimp}. 
The probability $p(P)$ of finding a certain eigenvalue $P$ can then be
calculated based on Eqs.\ (\ref{eq:psprobapprox}) and (\ref{eq:bsint}). When we
neglect all terms of order higher than linear in $\theta$ and keep in mind that
the first moments of $\ket{0}$ vanish, we will get the result $p(P)=|\psi(P)|^2$
in complete analogy to Eq.\ (\ref{eq:psprobapprox}).

Now we can continue with the expectation value of the position quadrature
$\hat{x}_M$, also neglecting all terms of order higher than linear in $\theta$,
which results in
\begin{equation}
 \textrm{E}^{(P)}[\hat{x}_M] = \theta \left\{ \textrm{Re}\left[X_w^{(P)}\right]
- \textrm{Im}\left[P_w^{(P)}\right]\right\}  . \label{eq:bsexpval}
\end{equation}
Furthermore we find 
\begin{equation}
  \textrm{Im}\left[ P_w^{(P)} \right] = \textrm{Im}\left[ 
\frac{\bra{P}\hat{P}_S\ket{\psi}}{\braket{P}{\psi}} \right]=0 ,
\end{equation}
since $\hat{P}_S$ is a Hermitian operator. Thus the expectation value Eq.\
(\ref{eq:bsexpval}) simplifies and is now equivalent to Eq.\
(\ref{eq:specexpval}) when we replace the interaction constant $g$ by the beam
splitter parameter $\theta$. The additional term $\hat{P}_S \hat{x}_M$ in the
beam splitter transformation Eq.\ (\ref{eq:bsint}), from which the imaginary
part of the weak value $P_w^{(P)}$ originates, has no influence on this
expectation value.
Hence a beam splitter interaction with a small parameter $\theta$ and a
specifically chosen postselection measurement can be used to obtain the real
part of the position quadrature weak value $X_w^{(P)}$. Based on this
information and the knowledge about the probability $p(P)$ of finding an
eigenvalue $P$ in the post\-selection measurement, the unknown state
$\ket{\psi}$ of the system mode $S$ can be reconstructed using exactly the
reconstruction relations derived in Eqs.\ (\ref{eq:recrelphase}) and
(\ref{eq:recrelabs}).

Note that the reconstruction relations for phase,  Eq.\ (\ref{eq:recrelphase}),
and modulus, Eq.\ (\ref{eq:recrelabs}), of a momentum representation rely on the
special choice of the weakly measured observable $\hat{A}_S=\hat{X}_S$ and the
postselection observable $\hat{B}_S=\hat{P}_S$ as well as on the form of the
interaction, Eq.\ (\ref{eq:bsint}). A different choice of observables and
interaction may thus lead to reconstruction relations suitable for
representations in terms of other degrees of freedom. This suggests that the
general idea behind this reconstruction scheme might be applicable to a broader
range of physical systems, not only light modes coupled by a beam splitter.

\section{Simulation \label{sec:sim}}

To see if the reconstruction scheme presented in the previous paragraphs is
really feasible, we will numerically simulate the setup shown in Fig.\
\ref{fig:qoimp} and investigate the various influences of relevant parameters.

\subsection{Influence of $\theta$}

We will first focus on the errors introduced by neglecting
higher orders of the beam splitter parameter $\theta$. Moreover, it is then
interesting to see how a statistical error in the measured observables due to a
finite number of measurement runs influences the reconstruction quality.
To give a first example, we choose the state
\begin{equation}
 \ket{\psi} = \mathcal{N}(\ket{\alpha=1} + \ket{\beta= 2\, \ee^{\ii \pi 4/5}} )
 \label{eq:exstate}
\end{equation}
of the system mode $S$, where $\mathcal{N}$ is a normalization constant and
$\ket{\alpha}$ and $\ket{\beta}$ are coherent states of light 
\cite{cs:schroedinger, cs:glauber1, *cs:glauber2}.
Note that this state in fact contains nontrivial phase and modulus dependencies
which allow us to simulate essential features of the reconstruction.

\begin{figure}
\subfigure[\ exact]{\includegraphics{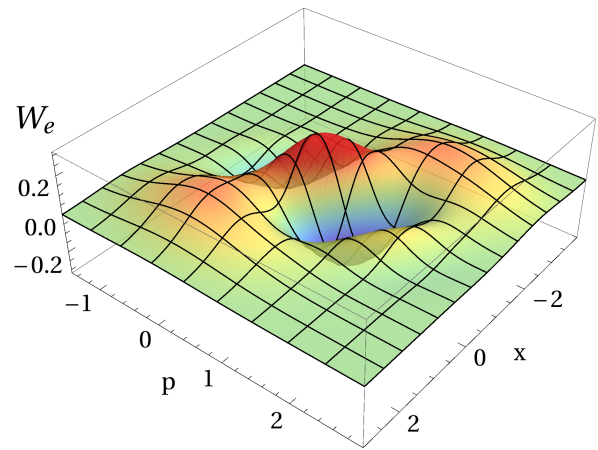}}
\subfigure[\ reconstruction 1: $\theta=0.05$, $\delta = 0.0003 $]{ 
\includegraphics{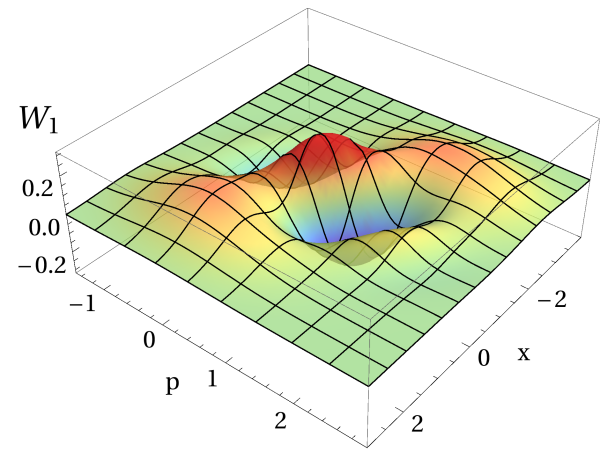}}
\subfigure[\ reconstruction 2: $\theta=\pi/4$, $\delta = 0.34$ 
\label{fig:5050}]{\includegraphics{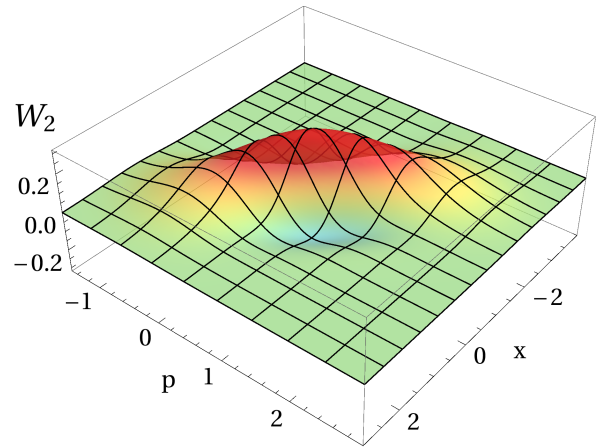}}
 \caption{(Color online) Error introduced by just keeping terms up to linear
order in $\theta$ in the weak-value expressions. Depicted are the corresponding
Wigner functions: $W_e$ of the exact state, Eq.\ (\ref{eq:exstate}), $W_1$ of
the reconstructed state $\ket{\psi_{rec}(\theta=0.05)}$ and $W_2$ of the
reconstructed state $\ket{\psi_{rec}(\theta=\pi/4)}$. The reconstruction error
$\delta$, Eq.\ (\ref{eq:recerror}), quantifies the total error. \label{fig:QOE}}
\end{figure}
First we have to decide how small $\theta$ shall be chosen to justify neglecting
all terms of order higher than linear in $\theta$. Therefore we calculate the
exact expectation value, Eq.\ (\ref{eq:exactexpval}), and the exact probability,
Eq.\ (\ref{eq:psprob}), by using the full beam splitter transformation, Eq.\
(\ref{eq:bsint}), applied to the system state $\ket{\psi}$, Eq.\
(\ref{eq:exstate}), and the meter state $\ket{\phi}=\ket{0}$. Next we insert the
results of these calculations into the reconstruction relations given in Eqs.\
(\ref{eq:recrelphase}) and (\ref{eq:recrelabs}). As these relations have been
derived disregarding all terms of order higher than linear in $\theta$, they
might be rough approximations, unless $\theta$ is small enough. To quantify the
error, we compute the reconstruction error
\begin{equation}
 \delta \equiv 1 - |\braket{\psi}{\psi_{rec}(\theta)}|^2 \label{eq:recerror}
\end{equation}
between the reconstructed state $\ket{\psi_{rec}(\theta)}$ and the exact state
$\ket{\psi}$.

In Fig.\ \ref{fig:QOE} we compare Wigner functions \cite{wigner1932, Hillery:wf, 
[{For a textbook introduction to Wigner functions see, for example, Ref.\
\cite{schleichQO}}][{}]du} of the reconstructed states to the Wigner function of
the exact state and we also show the corresponding reconstruction error, Eq.\
(\ref{eq:recerror}). It is clearly visible that the choice $\theta=0.05$ for the
beam splitter results in a small error introduced by retaining only terms of
order up to linear in $\theta$. Moreover, we also note that even for a 50:50
beam splitter [cf.\ Fig.\ \ref{fig:5050}] one can still reconstruct some
features of the state $\ket{\psi}$, Eq.\ (\ref{eq:exstate}), by using the
proposed scheme. However, negative parts of the Wigner functions are very
sensitive to a badly chosen parameter $\theta$.

Remarkably for a 50:50 beam splitter, i.e., $\theta=\pi/4$, the setup depicted in
Fig.\ \ref{fig:qoimp} is nothing else than an eight-port interferometer (see,
for example, \cite{schleichQO}). In this case the reconstruction error $\delta$
is large for our way of processing the measured data. Yet the Husimi-Kano
$Q$ function \cite{Husimi,kano} may be obtained directly as the joint probability
distribution of such an eight-port interferometer \cite{walkercarroll}. However,
it is quite complicated to extract the underlying quantum state from a
$Q$ function. On the other hand, if $\theta$ is small we can directly obtain the
absolute value and the phase of an unknown state by using the simple relations
Eqs.\ (\ref{eq:recrelphase}) and (\ref{eq:recrelabs}). At this point it is
interesting to see how in two different measurement regimes two very different
ways of processing measured data each lead to the full state information.

\subsection{Statistical error}

Up to now we have dealt with exact expectation values and full probability
distributions, which in a real experiment are accessible only in the limit of an
infinite number $N$ of measurement runs. To account for the influence of the
statistical error due to a finite $N$, we also perform a Monte Carlo simulation
for the optical setup shown in Fig.\ \ref{fig:qoimp} with the state of
Eq.\ (\ref{eq:exstate}). For this simulation, as for any real experiment, we
furthermore have to choose a certain binning $\Delta P$ \cite{[{The bin size we
choose throughout this article is an empirical value that seems to be
appropriate for the performed simulations. As the transformation of simulated
data invokes only a square root, Eq.\ (\ref{eq:recrelabs}), and an integration,
Eq.\ (\ref{eq:recrelphase}), the results are quite stable to a variation of the
bin size. Of course for a finite amount of measurement runs the bin size has to
unite two opposing goals: (i) a small bin size allows us to perfectly map the
shape of the probability distribution whereas (ii) it should be chosen as large
as possible to increase the number of counts per bin, thus minimizing the
statistical error. Therefore, an optimization is an appealing challenge which
is closely connected to the methods of }][{}]Bellini} of the momentum quadrature
values $P$. Hence we obtain approximated probability distributions for the
$P$ quadratures and approximated values for the postselected expectation values
Eq.\ (\ref{eq:bsexpval}). These estimates for probability and expectation value
are then inserted into the reconstruction relations Eqs.\
(\ref{eq:recrelphase}) and (\ref{eq:recrelabs}).

\begin{figure}[ht!]
\includegraphics{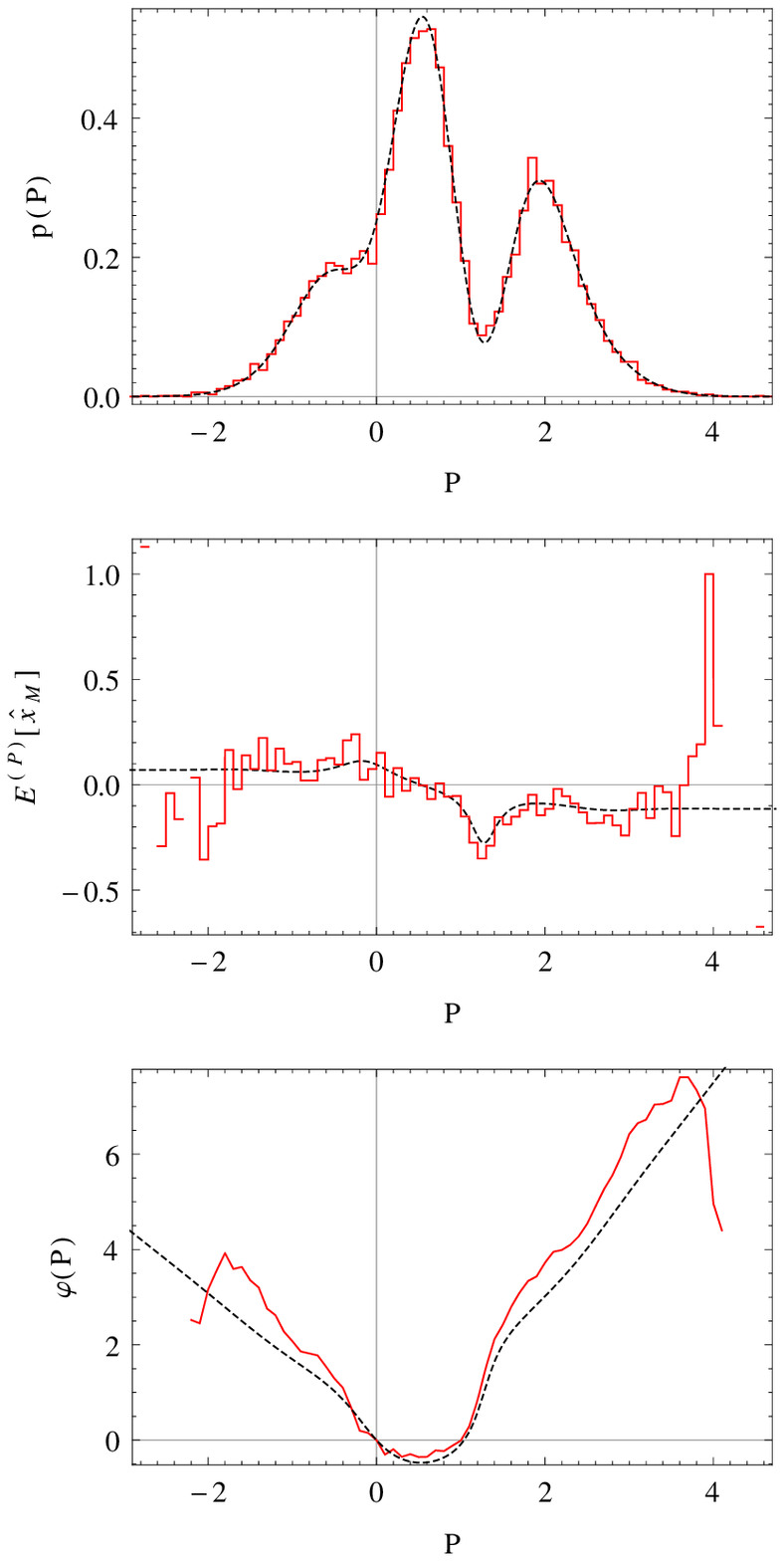}
 \caption{(Color online) Monte Carlo simulated (red, solid) and exact (black,
dashed) values of probability $p(P)$, expectation value
$\textrm{E}^{(P)}[\hat{x}_M]$, and phase $\varphi(P)$ as a function of the
momentum quadrature $P$ for the state in Eq.\ (\ref{eq:exstate}). Depicted are
results for $N=10^4$ measurement runs for a beam splitter parameter
$\theta=0.05$ and a bin width $\Delta P=0.1$. The reconstruction error is
$\delta=0.08$. One can see that the statistics is good enough to reconstruct the
modulus $|\psi(P)|=\sqrt{p(P)}$. However, the postselected expectation value
shows considerable deviations. This is even more pronounced in those regions
where we have a small probability to find a certain value of the momentum
quadrature. As a consequence the reconstructed phase soon differs from the exact
value. \vspace*{0.5cm} \label{fig:recresults10e4}}
\end{figure}
\begin{figure}[ht!]
\includegraphics{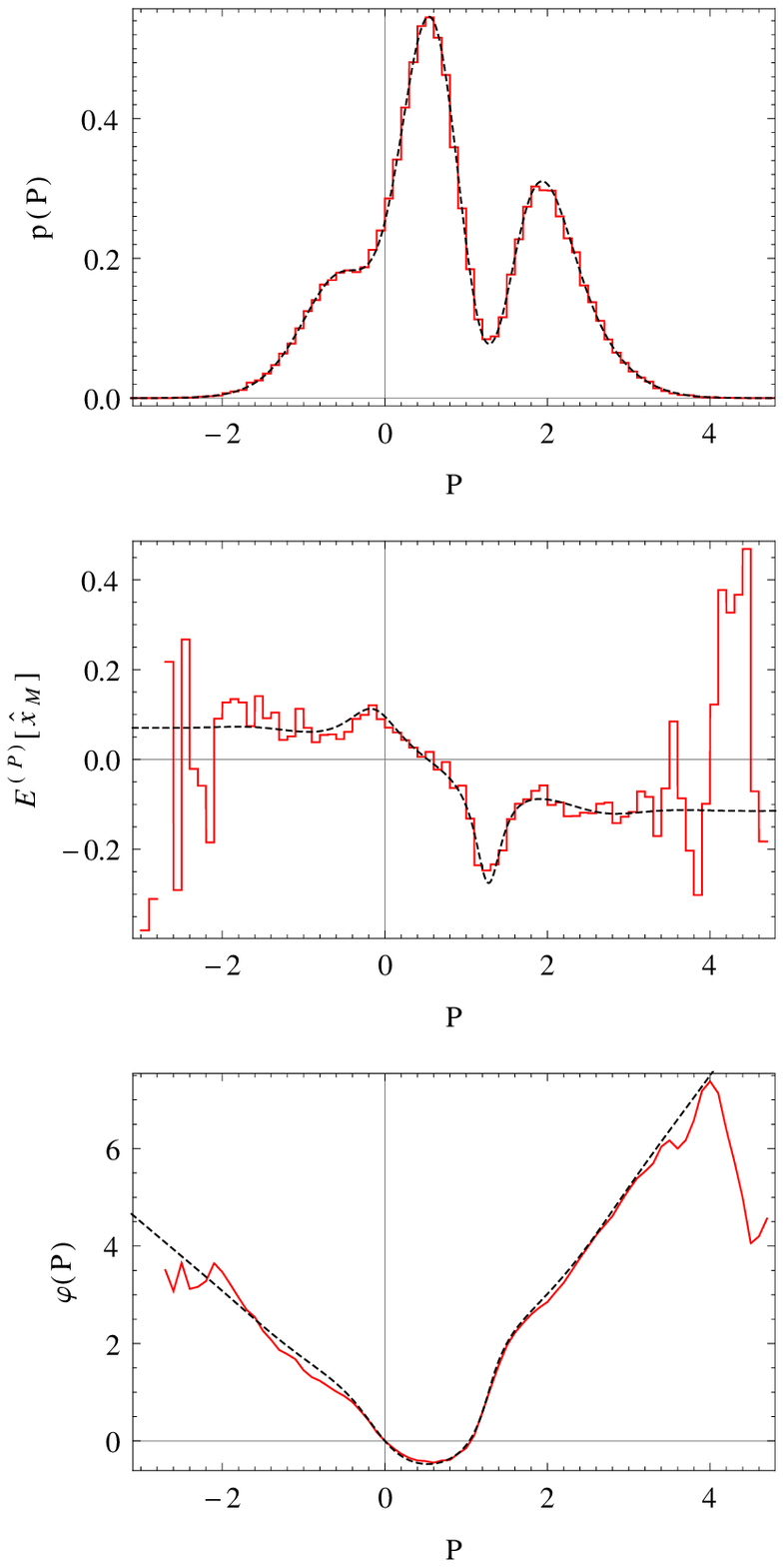}
 \caption{(Color online) Monte Carlo simulated (red, solid) and exact (black,
dashed) values of probability $p(P)$, expectation value
$\textrm{E}^{(P)}[\hat{x}_M]$, and phase $\varphi(P)$ as a function of the
momentum quadrature $P$ for the state in Eq.\ (\ref{eq:exstate}). Depicted are
results for $N=10^5$ measurement runs for a beam splitter parameter
$\theta=0.05$ and a bin width $\Delta P=0.1$. The reconstruction error is
$\delta=0.008$. In comparison with Fig.\ \ref{fig:recresults10e4} the statistics
here is much better. The postselected data now show stronger fluctuations only
in those regions where the probability $p(P)$ is small. Hence the reconstructed
phase deviates considerably only when the modulus has already become
exponentially small. \vspace*{0.7cm} \label{fig:recresults10e5}} 
\end{figure}
Figures \ref{fig:recresults10e4} and \ref{fig:recresults10e5} represent two 
typical measurements for a properly chosen beam splitter parameter
$\theta=0.05$. The number of measurement runs is $N=10^4$ and $N=10^5$,
respectively. 
We show the pure measurement data, namely, approximated probabilities and
approximated expectation values, as well as the resulting reconstructed phase.
The steps in the probabilities and expectation values as well as the edges in
the reconstructed phases are an artifact of the finite bin width $\Delta P$.
As expected, increasing the number of measurement runs $N$ improves the
reconstruction quality.

Furthermore, we notice that in general those parts of the reconstructed
phases located in the vicinity of $P=0$ are in better agreement with the exact
phase than outer parts situated at larger values of $P$.
There are two reasons for this.
First, according to Eq.\ (\ref{eq:recrelphase}), we start the integration of the
postselected expectation value at $P=0$. Therefore, the reconstructed and exact
phases agree perfectly at this point. Then we need more and more expectation
values to reconstruct the phase for increasing values of $P$. Hence the
statistical errors involved in every single expectation value sum up. 
Second, the statistical error of a single expectation value increases when the
corresponding probability of obtaining the outcome $P$ decreases.
This is most clearly visible from the outer parts of the estimated expectation
values in Figs.\ \ref{fig:recresults10e4} and \ref{fig:recresults10e5}, which
tend to fluctuate more strongly around the exact value as the probability
$p(P)$ gets smaller.

Due to these intricacies, the reconstructed phase is more sensitive to
statistical errors than the reconstructed modulus. Even for $N=10^5$ measurement
runs (see Fig.\ \ref{fig:recresults10e5}) the reconstructed phase does not match
the exact value as nicely as does the reconstructed modulus. For $N=10^4$ (see
Fig.\ \ref{fig:recresults10e4}) this contrast is even more pronounced. As in
other reconstruction schemes \cite{leonhardt, paris2004quantum, freyberger}, it
is harder to reconstruct the phase than the absolute value of the quantum state.

However, in the scheme discussed here this sensitivity of the phase can be fully
traced back to the measured data: The set of postselected expectation values
needs to be known with good statistical confidence and on the whole $P$ axis.
This also means that quantum states $\ket{\psi}$ with regions in the quadrature
distribution $|\psi(P)|^2$ that are nonzero and separated by an intermediate
region where the quadrature distribution is equal to zero cannot be
reconstructed easily. In these gaps we will never find a postselected
expectation value and hence we will lose the phase relation between parts of
the quadrature representation $\psi(P)$ separated by these gaps.
To a certain degree this can be seen already in Figs.\ \ref{fig:recresults10e4}
and \ref{fig:recresults10e5}. For some values of $P$ we are missing the
corresponding expectation values (see, e.g., $P=-2.75$ in Fig.\
\ref{fig:recresults10e5}), and hence we lose track of the phase. However, as
this happens here in a region where $|\psi(P)|$ stays exponentially small, it is
not very problematic and the reconstruction error $\delta$ is still small.

\subsection{Counterexample}

An explicit example of a state showing a distinct gap in the probability
distribution $|\psi(P)|^2$ is the state
\begin{equation}
 \ket{\psi} = \mathcal{N} \left( \ket{\alpha=2 \ii} - \ket{\beta= - 2 \ii}
 \right), 
\label{eq:cestate}
\end{equation}
where $\mathcal{N}$ again marks the normalization constant for a superposition
of two coherent states $\ket{\alpha}$ and $\ket{\beta}$.
\begin{figure}[ht!]
\includegraphics{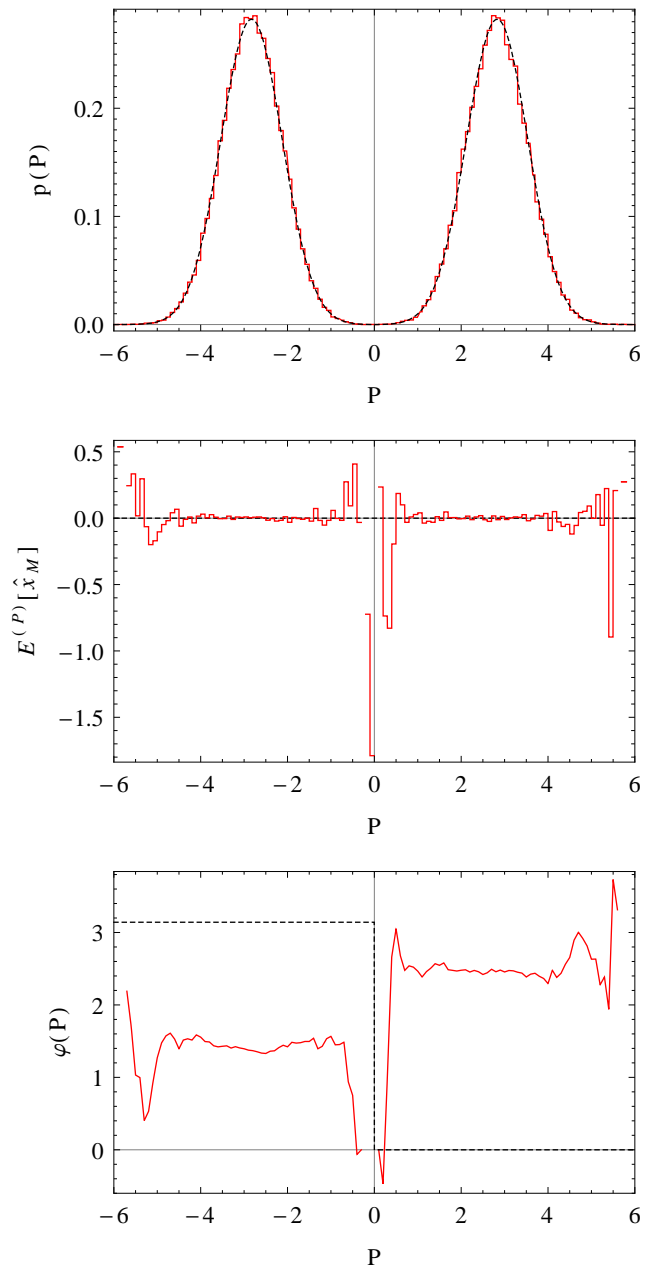}
 \caption{(Color online) Monte Carlo simulated (red, solid) and exact (black,
dashed) values of probability $p(P)$, expectation value 
$\textrm{E}^{(P)}[\hat{x}_M]$, and phase $\varphi(P)$ as a function of the 
momentum quadrature $P$ for the state in Eq.\ (\ref{eq:cestate}). Depicted are
results for $N=10^5$ measurement runs for a beam splitter parameter
$\theta=0.05$ and a bin width $\Delta P=0.1$. The small probability at the
origin causes gaps in the simulated expectation values. Hence we lose track of
the phase at the origin and can only guess the relative phase between the two
regions left and right of the origin. Therefore direct reconstruction of this
state fails. \label{fig:recresultsce}} 
\end{figure}
In Fig.\ \ref{fig:recresultsce} we show the corresponding Monte Carlo simulated
data in comparison to the exact results. The probability $|\psi(P)|^2$ around the
origin is so small that there are values of $P$ which never appear on our meter
HDS in all of the $N=10^5$ simulated measurements. It is exactly those values of
$P$ where we can find no estimate for the postselected expectation value.
Therefore, the reconstructed expectation value is not continuous around the
origin and we cannot use Eq.\ (\ref{eq:recrelphase}) to reconstruct the phase
on the complete $P$ axis.

However, as shown in Fig.\ \ref{fig:recresultsce}, we can integrate the
reconstructed expectation values in those regions where  $|\psi(P)|^2$ is large
enough and reconstruct a phase. To this end we have to choose two different
starting points of the integration Eq.\ (\ref{eq:recrelphase}), each lying
within one of these regions. This means we explicitly pick a certain ``global''
phase in each of the two regions. Hence we introduce an arbitrary relative phase
shift between the two regions.

Now let us compare this to the exact result also shown in Fig.\
\ref{fig:recresultsce}. The exact value of the phase $\varphi(P)$ jumps by $\pi$
at the origin and is otherwise constant. The two parts of the reconstructed
phase show this constant behavior in the corresponding regions where the
influence of the statistical error is small, i.e., where $|\psi(P)|^2$ is
reasonably large. The difference between these two constant values is exactly
the relative phase shift introduced by simply choosing a ``global'' phase in
each region separately. Therefore, without any additional measurements we cannot
determine this relative phase shift. Consequently for this example
reconstruction fails. Fortunately, gaps in the probability distribution are
visible from the measured data and therefore indicate whether we can rely on the
reconstructed state or not.

A possible workaround is the reconstruction of a unitarily transformed state,
not showing any gaps. One example would be to measure suitable rotated quadratures
on the homodyne detectors. This is equivalent to implementing a unitary
transformation rotating the Wigner function of the input state. But this
procedure will not dispose of gaps in a state with a rotationally invariant
Wigner function, such as, for example, a Fock state. So there is no general recipe to
find a unitary transformation that is easily implementable and removes the gaps.

\section{Conclusion}

In this paper we have shown how to use weak measurements to reconstruct a
continuous-variable state in quantum optics. The modulus of a suitably
chosen representation can be reconstructed from projective measurements on the
system itself, while the corresponding phase has to be extracted from a complete
set of postselected expectation values measured on a weakly coupled meter
device. An appealing facet of the presented reconstruction scheme is the
simplicity of the reconstruction relations which turn measured data into quantum
state information. Moreover, this measurement concept can be realized with basic
elements of linear optics and homodyne detection. However, the sorting required
for the postselection of data poses additional problems. The reconstruction of
the phase is not straightforward for states with gaps in the probability
distribution which determines the sorting. A generalization of the presented
scheme to mixed states is desirable, but based on the presented reconstruction
relations not straightforward.

\vspace*{1cm}

%\bibliography{qsrwv}


\end{document}